# A pseudopotential multiphase lattice Boltzmann model based on high-order difference

Zhangrong Qin, Wanling Zhao, Yanyan Chen, Chaoying Zhang[*], Binghai Wen[*]

Guangxi Key Lab of Multi-Source Information Mining & Security, Guangxi Normal University, Guilin 541004, China

The hyperbolic tangent function is usually used as a reliable approximation of the equilibrium density distributions of a system with phase transitions. However, analyzing the accuracies of the numerical derivatives, we find that its numerical derivatives computed by central difference method (CDM) may deviate significantly from its analytical solutions, while those computed by high-order difference method (HDM) can agree very well. Therefore, we introduce HDM to evaluate the interparticle interactions instead of popular CDM, and propose a pseudopotential multiphase lattice Boltzmann model based on high-order difference method. The present model not only retains the advantages of the pseudopotential model, such as easy implementation, high efficiency, full parallelism and so on, but also achieves higher accuracies. To verify the performances of this model, several multiphase flow simulations are conducted, including both stationary and dynamic situations. Firstly, full thermodynamic consistencies for the popular equations of state have been achieved in large temperature range and at large density ratio, without any combining interaction and any additional adjustable parameter of interaction. Secondly, with high-order difference, either the interparticle interaction proposed by Shan-Chen or by Zhang-Chen can equally depict the phase transitions of the fluids with all selected equations of the state. These numerical

---

[*] Corresponding authors.

E-mail address: zhangcy@gxnu.edu.cn (C. Zhang), oceanwen@gxnu.edu.cn (B. Wen)

**agreements based on HDM are consistent to the theoretical analysis that the two models are mathematically identical. Thirdly, the present model is stable and accurate at a wider temperature range. Lastly, our newly proposed model can be easily and reliably applied to various practical simulations and expected to obtain some more interesting results.**

# 1. Introduction

Due to its numerous advantages, including clear physical backgrounds, simple algorithm, high efficiency and full parallelization, lattice Boltzmann method (LBM) has been successfully applied in modeling complex fluid systems involving interfacial dynamics and phase transitions [1, 2]. In past two decades, a considerable number of efforts have been devoted to developing multiphase lattice Boltzmann models, and three main kinds of models have been proposed, i.e., color model proposed by Gunstensen et al. [3], pseudopotential methods by Shan and Chen [4, 5] and free energy model by Swift et al. [6]. Among them, the pseudopotential model has been popularly applied due to its easy implementation and high efficiency in multiphase flow simulations [7, 8]. In the original pseudopotential model, the interparticle force is mimicked by a so called "effective mass", $\psi(\bm{x})$, which is usually set to $\psi(\bm{x}) = 1 - \exp(-\rho/\rho_0)$ empirically, where $\rho_0$ is a reference density. Although this model is able to reproduce phase transition phenomena, it is thermodynamic inconsistent, and cannot be applied to those fluids with the usual equations of state or applied to large density ration systems. By incorporating different equations of state (EOS) into the pseudopotential model, Yuan and Schaefer [9] extended its application range, and achieved higher density ratios. However, as the interparticle force form remains unchanged, it is still thermodynamic inconsistent slightly. Furthermore, the applying range of the temperature of the improved model is still small, and thus the density ratio is limited. To obtain thermodynamic consistency and achieve larger density ratios, Kupershtokh et al. [10, 11] proposed a new pseudopotential model by integrating reduced equations of state, in which the interparticle interaction is evaluated by combining the general effective mass method [4] and the method based on potential function [12]. Consequently, the simulation stability is greatly improved, the temperature range is extended, and the thermodynamic consistency is also ensured numerically. Subsequently, Gong et al. [13] and Hu et al. [14] also proposed pseudopotential models like Kupershtokh's successively, and good results were again retrieved as expected. In Hu's model, a method of modifying equations of state was proposed in terms of the Maxwell construction theory, and the simulation stability was greatly enhanced with the modified equations of state. Due to their excellent performances, the three hybrid interaction models have been widely used in multiphase flow simulations [15-18]. For constructing the pseudopotential multiphase lattice Boltzmann model, from the usual views, two problems must be addressed: obtaining a correct interaction with clear physical concept and solid theory foundation, and developing an effective method to incorporate the interaction into the basic lattice Boltzmann equation. In recent years, many works have been done to obtain better pseudopotential interactions, and three main forms with good performances have been developed, including "effective mass" interactions [4], interactions with potential forms [12], and the hybrid interactions by combining the former two forms with weight coefficient method [10, 13, 14]. Among them, the hybrid interactions have been popularly used owing to their high accuracies. As the methods of incorporating the interaction into the lattice Boltzmann equation, so called forcing technologies, many variants of the method have been developed [4, 10, 19-22]. These forcing technologies can be all used to incorporate the interactions into the basic lattice Boltzmann equations with different degrees of approximation. It is worth to mentioning that the technology proposed by Kupershtokh et al. is simple, accurate and independent of the relaxation time, thus it becomes popular recently. Reviewing the developments of the pseudopotential multiphase lattice Boltzmann models, no matter the interparticle interaction forms or the forcing technologies have been achieved considerable progresses, it seems as if two of the most important problems have already resolved. Deeply analyzing the existing interactions forms shows that, both the effective mass interaction and the interaction with potential form have numerous successful applications, and they really are identical mathematically. From the literatures [10, 13, 14], neither of these two interaction forms can well describe the phase transitions of the usually used equations of state, however their linear combination with an adjustable coefficient gives better descriptions. This means that the linear combination of two mathematically identical interactions provides new physical results, which is hardly understood. On the other hand, in order to obtain the optimal results, the adjustable parameter in the hybrid interaction must take different values for different equations of state, thus this interaction form may have artificial arbitrariness [10, 13]. Furthermore, seeing the literatures [10, 13, 14], the adjustable parameter needs to be negative or larger than 1 in order to obtain the optimal solutions in these cases; this parameter may lose its meaning of weight. Therefore, we think that there should be a third issue for constructing the pseudopotential multiphase lattice Boltzmann models, which may have not been attracted much attentions before, e.g. for a specific interaction, how can we numerically evaluate the interaction with the required accuracies? As far as we know, this third problem has seldom been investigated in the published literatures. In this paper, we will investigate the numerical methods for calculating the interactions, and reveal the importance of the numerical accuracies of the interactions. Then, we introduce a new numerical scheme to evaluate the interparticle interactions and construct an effective pseudopotential multiphase lattice Boltzmann model. Finally, several testing cases are conducted to demonstrate the performances of the proposed model.

The rest of this paper is organized as follows. Firstly, a brief introduction is given to the pseudopotential multiphase lattice Boltzmann model. Secondly, the central difference and the high-order difference methods are investigated. Thirdly, a pseudopotential multiphase lattice Boltzmann model based on high-order difference is constructed. Fourthly, the performances of the present model are investigated by using the multiphase flows with several usual equations of state, such as van der Waals EOS (VDW EOS), Peng-Robinson EOS (PR EOS), Carnahan-Starling EOS (CS EOS) and Redlich-Kwong EOS (RK EOS). Finally, a brief conclusion is drawn.

## 2. Pseudopotential models

Originating from the cellular automaton concept and kinetic theory, the intrinsic mesoscopic properties make lattice Boltzmann method (LBM) outstanding in modeling complex fluid systems involving interfacial dynamics[23-25] and phase transitions [7, 8]. In basic approach of LBM, a particle distribution function is used to

depict the behavior of the multiphase flows, which obeys the following lattice Boltzmann equation

$$f_i(\boldsymbol{x}+\boldsymbol{e}_i\delta t, t+\delta t) - f_i(\boldsymbol{x},t) = -\frac{1}{\tau}[f_i(\boldsymbol{x},t) - f_i^{(eq)}(\boldsymbol{x},t)] + F_i, \tag{1}$$

where $f_i(\boldsymbol{x},t)$ is the particle distribution function at lattice site $\boldsymbol{x}$ and time $t$, $\delta t$ is the time step, which is usually taken to be unit, $\boldsymbol{e}_i$ with $i=0,...,N$ is the discrete speed, $\tau$ is the relaxation time, and $f_i^{(eq)}$ is the equilibrium density distribution function, it can be represented as (D2Q9 model)

$$f_i^{(eq)}(\boldsymbol{x},t) = \rho\omega_i[1 + 3(\boldsymbol{e}_i\cdot\boldsymbol{u}) + \frac{9}{2}(\boldsymbol{e}_i\cdot\boldsymbol{u})^2 - \frac{3}{2}u^2], \tag{2}$$

where $\omega_i$ is the weight coefficient, and $\boldsymbol{u}$ is the fluid velocity. The fluid density $\rho$ and the velocity $\boldsymbol{u}$ (in the absence of body fore) can be calculated by

$$\rho = \sum_{i=0}^{N} f_i, \quad \rho\boldsymbol{u} = \sum_{i=0}^{N} \boldsymbol{e}_i f_i, \tag{3}$$

In pseudopotential models, the body force term $F_i$ is the discrete version of the interparticle interaction. A widely used interparticle interaction is proposed by Shan-Chen [4], which is so called "the effective mass based interaction" and can be written as

$$\boldsymbol{F}^{SC}(\boldsymbol{x},t) = -G\psi(\rho(\boldsymbol{x},t))\nabla\psi(\rho(\boldsymbol{x},t)), \tag{4}$$

where $\psi(\rho(\boldsymbol{x},t))$ is the "effective mass", and $G$ is the interaction strength. In the early applications, "effective mass" is usually taken the simple form $\psi(\rho) = 1 - \exp(-\rho/\rho_0)$, and thus the model can't be directly applied to the fluids with usual equations of state. With equation (4), the corresponding equation of state is

$$p = c_s^2\rho + \frac{G}{2}\psi^2(\rho), \tag{5}$$

In order to extend the pseudopotential model for including the fluids with the usual equations of state, Yuan and Schaefer [9] incorporated the equations of state into LBM by introducing the following relation between the "effective mass" and the equation of state

$$\psi(\rho) = \sqrt{\frac{2(p - \rho c_s^2)}{G}}, \tag{6}$$

where $c_s^2$ is the "sound speed", and takes the value $c_s^2 = \frac{1}{3}$ in D2Q9 model.

Another popular interparticle interaction is presented by Zhang-Chen [12], which is so called "the interaction based on potential function" and can be written as

$$\boldsymbol{F}^{ZC}(\boldsymbol{x}, t) = -\nabla U(\boldsymbol{x}, t), \tag{7}$$

where $U(\boldsymbol{x}, t)$ takes the following form

$$U(\boldsymbol{x}, t) = p - \rho c_s^2. \tag{8}$$

Recently, a third popular interaction was proposed [10, 13, 14], which was obtained by linearly combining both the former two with a weighted coefficient.

$$\boldsymbol{F}(\boldsymbol{x}, t) = A\boldsymbol{F}^{ZC}(\boldsymbol{x}, t) + (1 - 2A)\boldsymbol{F}^{SC}(\boldsymbol{x}, t), \tag{9}$$

where $A$ is the weight coefficient.

To obtain the force term $F_i$ from the macroscopic interaction $\boldsymbol{F}$, the popular forcing technology proposed by Kupershtokh et al.[10] is used in this paper due to its particular advances. It can be described as

$$F_i = f_i^{(eq)}(\rho, \boldsymbol{u} + \Delta \boldsymbol{u}) - f_i^{(eq)}(\rho, \boldsymbol{u}), \tag{10}$$

where $\Delta \boldsymbol{u} = \frac{\boldsymbol{F}(x,t)}{\rho} \delta t$. What is to say that the body force term $F_i$ is simply equal to a difference of equilibrium distribution functions corresponding to the mass velocity after and before the action of a force during a time step at constant density $\rho$. And in this case, instead of $\boldsymbol{u}$, the real fluid velocity should be taken at half time step

$$\rho \boldsymbol{v} = \sum_{i=0}^{N} \boldsymbol{e}_i f_i + \frac{\boldsymbol{F}}{2} \delta t, \tag{11}$$

Using the Chapman-Enskog expansion, Eq. (1) can recover the following macroscopic Navier-Stokes equations ignoring high-order additional terms [26]

$$\frac{\partial \rho}{\partial t} + \nabla \cdot (\rho \boldsymbol{v}) = 0, \tag{12}$$

$$\frac{\partial(\rho \mathbf{v})}{\partial t} + \nabla \cdot (\rho \mathbf{v}\mathbf{v}) = -\nabla p_0 + \mathbf{F} + \nabla \cdot [\rho \nu (\nabla \mathbf{v} + (\nabla \mathbf{v})^T)], \tag{13}$$

where $p_0 = c_s^2 \rho$ is the ideal fluid pressure, and $\nu = (2\tau - 1)/6$ is the viscosity coefficient.

## 3. Central difference and high-order difference method

Analysis of the above pseudopotential models shows that, almost all the interparticle interactions include the gradients of the macroscopic density or effective mass. This means that the evaluations of gradients take an important part in the modeling of multiphase flows. Up to now, the gradients in interactions are usually evaluated by using the central difference method [10, 13, 14, 27] due to its simplicity, high efficiency and reliable accuracy for smooth and slow varying functions. Few literatures have addressed the resulting errors of the central difference method in simulating multiphase flows. In general cases, the central difference method is accurate enough so that its accuracy is seldom in doubt. It seems as if this problem may be ignored in the studies of the multiphase flows by using the pseudopotential models. It is well known that the numerical derivatives may suffer from ill posed problems, and then the numerical gradients by the central difference may also deviate significantly from the exact values. Thus, we will firstly investigate the applications of the central difference in some specific cases, which are closer to that of the real multiphase flows, and then proposed a numerical scheme for evaluating the gradients with better accuracies.

3.1 Central difference method

Consider a continuous and derivable function $y = f(x) \in [a, b]$, $a \leq b$, and divide the interval $[a, b]$ into n subintervals with equal spacing $h = \dfrac{b-a}{n}$ and separated by $x_i (i = 0, 1, \cdots, n-1, n)$. In terms of CDM, the numerical derivative at $x_i$

can be written as

$$y'(x_i) = \frac{y(x_{i+1}) - y(x_{i-1})}{2h}, \tag{14}$$

It is obvious that the derivate at node $x_i$ is determined by the function values of $f(x_{i-1})$ and $f(x_{i+1})$ at its neighbors, and independent of $f(x_i)$ at its own site. In the real multiphase flows simulations, instead of CDM, a weighted central difference scheme is usually applied to evaluate the gradients. In the D2Q9 model, it takes the following form for the numerical gradient of density [27]

$$\frac{\partial \rho(\mathbf{x})}{\partial x} = \sum_{i=0}^{8} \omega_i \rho(\mathbf{x}+\mathbf{e}_i) e_{ix}, \quad \frac{\partial \rho(\mathbf{x})}{\partial y} = \sum_{i=0}^{8} \omega_i \rho(\mathbf{x}+\mathbf{e}_i) e_{iy}, \tag{15}$$

where $\omega_i$ is the weighting coefficient which is the same as that in Eq. (2). For a system with phase transitions, its equilibrium distribution of density can be usually approximated with a hyperbolic tangent function. To research the performances of CDM in the multiphase fluid simulations, we start from a hyperbolic tangent function

$$\rho(x) = 1 + Tanh(\frac{2x}{W}), \tag{16}$$

where $W$ indicates the width of the interface between the two phases. The first derivatives of density calculated with CDM, in which $h=1$ and $W=3,5$ are shown on Fig. 1, and the corresponding exact results are also plotted for comparison. In order to further know the accuracies of CDM, we evaluate the absolute errors in term of the relation

$$E_r = \left| \rho'_i - \rho'_{exact} \right|, \tag{17}$$

here $\rho'_i$ is the numerical result and $\rho'_{exact}$ the analytical value. The calculated results are given in Fig. 2. For investigating the errors of CDM on different widths of the interfaces between two phases, we also calculate the standard error by the following formula

$$Err = \sqrt{\frac{1}{n}\sum_{i=1}^{n}(\rho'_i - \rho'_{exact})^2}, \tag{18}$$

The evaluated results for first derivative of density are shown in Fig. 3.

From Fig. 1, we can clearly see that the numerical results by CDM deviate significantly from the analytical ones, and the numerical results with W=5 are more accurate than those with W=3. These deviations may affect the simulation accuracies, and even give incorrect results. Figs. 2 and 3 show that, with the decrease of the width W, the deviations increase steeply. While in many usual multiphase simulations, the widths of the interfaces between two phases are about W=1, the results from the errors of the central difference scheme may be fatal. Therefore, another difference method with high order accuracies is necessary to be utilized for well describing the behavior of the multiphase flows, especially at low relative temperatures (far below the critical temperature).

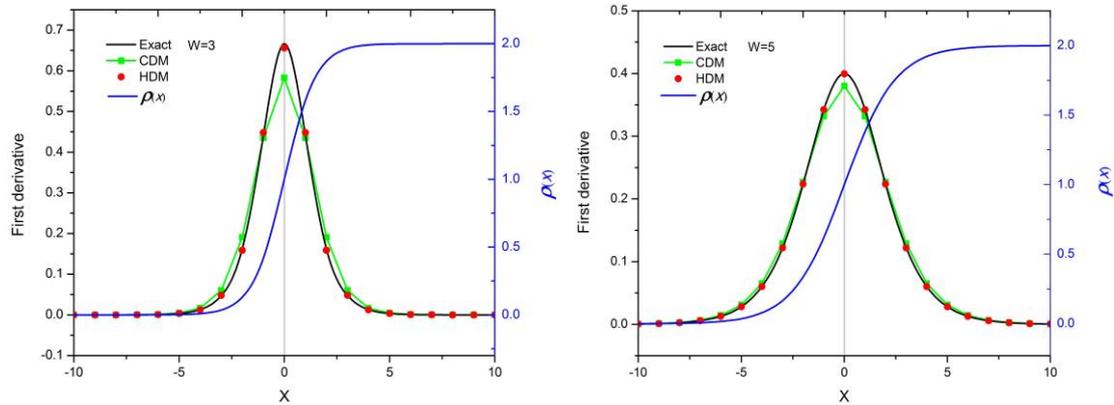

Fig. 1. The first derivatives evaluated with CDM and HDM for Eq. (16)

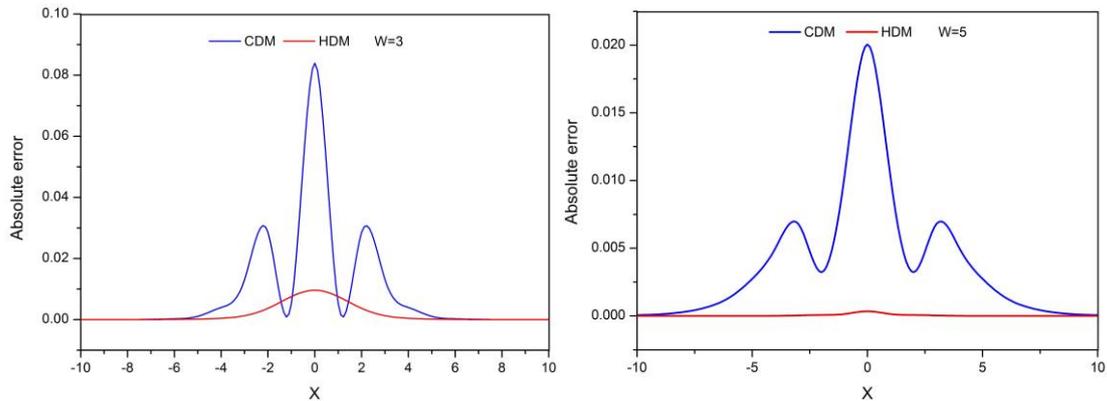

Fig. 2. The absolute errors of the first derivatives from CDM and HDM for Eq. (16).

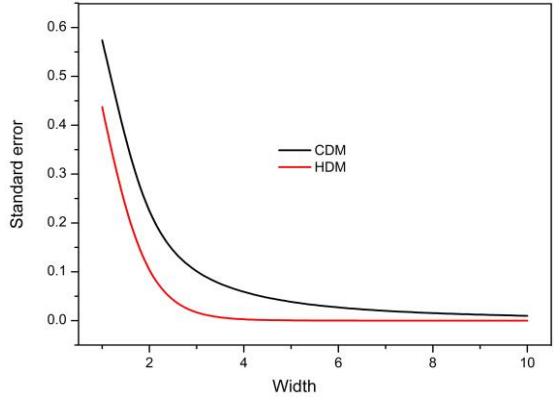

Fig. 3. The standard errors vs width for the first derivatives from CDM and HDM.

3.2 High-order difference method

CDM determines the derivative on some site by only using the data of its neighboring nodes; the evaluation is pointwise and thus has the advantages of implementation easy and high efficiency. Therefore, CDM are very popular in the pseudopotential models. However, without its own information, the derivative by CDM is merely an average on its neighboring nodes, and its precision is not satisfactory in the steep interface between gas and liquid phases in the multiphase flow problems.

A high-order difference scheme is widely used in the finite difference methods. Its high accuracy has been confirmed by many of the successful applications with the finite difference method. Thus, it is naturally to introduce a high-order difference scheme to evaluate the gradients in the pseudopotential models.

Different from the CDM, several derivatives are related to multiple differences in HDM; and this makes the derivative at some node not only depending on the multiple function values at its neighboring nodes, but also that at its own point. Without loss of generality, consider a continuous and derivable function $u(x) \in [a, b]$, $a \leq b$. The interval $[a, b]$ is then divided into n subintervals with equal spacing, $x_i = a + (i-1)h$, $h = (b-a)/(n-1)$, $i = 1, 2, \cdots, n$, and let $u_i = u(x_i)$, $u_i' = u'(x_i)$, $u_i^{(m)} = u^{(m)}(x_i)$, $\cdots$. Using the method of undetermined coefficients, we can write the

following relation between the multiple derivatives and the multiple differences

$$\beta u'_{i-2} + \alpha u'_{i-1} + u'_i + \alpha u'_{i+1} + \beta u'_{i+2} = c\frac{u_{i+3} - u_{i-3}}{6h} + b\frac{u_{i+2} - u_{i-2}}{4h} + a\frac{u_{i+1} - u_{i-1}}{2h}, \quad (19)$$

where $\alpha, \beta, a, b$ and $c$ are undetermined coefficients. Considering the requirements of both accuracy and simplicity, we take $\beta = 0$ and $\alpha = 1/3$, and then the following simplified relation can be obtained

$$\frac{1}{3}u'_{i-1} + u'_i + \frac{1}{3}u'_{i+1} = \frac{1}{9}\frac{u_{i+2} - u_{i-2}}{4h} + \frac{14}{9}\frac{u_{i+1} - u_{i-1}}{2h} \quad (20)$$

Eq. (20) is a typical tridiagonal equation; it can be solved with the chasing method. In order to show the performances of the HDM, the similar calculations on the density of Eq. (16) with HDM have been also performed. The corresponding results are also plotted in Figs. 1 and 2. The standard error of HDM vs. the width is given in Fig.3. Figs. 1 and 2 clearly show that the accuracies of HDM are much higher than those of CDM. With the decrease of the interface width between two phases, the standard error from HDM also increases, but its error level is much lower than that of CDM. Finally, we draw a conclusion that, instead of CDM, the numerical gradients in multiphase simulations with HDM may be more accurate.

## 4. Pseudopotential multiphase lattice Boltzmann model based on high-order difference

As discussed above, for constructing a pseudopotential model based on high-order difference, not two but three crucial problems must be addressed: (1) deriving an appropriate interparticle interaction, which has a solid theory foundation and a clearly physical background, (2) incorporating the interaction into the basic lattice Boltzmann equation, and (3) selecting an optimum method to evaluate the interparticle interaction.

As for the interparticle interactions, the earliest one proposed by Shan-Chen, later improved by Yuan and Schaefer, is a phenomenological force, which is dependent on the "effective mass" and has an explicit expression of Eq. (4). This interaction has been already successfully used in many applications in micro/nanoscience. Another interaction proposed by Zhang-Chen is based on potential form, and it can be described as Eq. (7). This force has also many successful applications in multiphase flow simulations. Introducing a special function by Eq. (6)

$$\Phi(\rho,T) = \begin{cases} \sqrt{p - \rho c_s^2}, & if\ p - \rho c_s^2 \geq 0 \\ \sqrt{\rho c_s^2 - p}, & if\ p - \rho c_s^2 < 0 \end{cases}, \quad (21)$$

Eqs. (4) and (7) can be rewritten as

$$\boldsymbol{F}(\boldsymbol{x},t) = \begin{cases} -2\Phi(\rho,T)\nabla\Phi(\rho,T), & if\ p - c_s^2\rho \geq 0 \\ 2\Phi(\rho,T)\nabla\Phi(\rho,T), & if\ p - c_s^2\rho < 0 \end{cases}, \quad (22)$$

and

$$\boldsymbol{F}(\boldsymbol{x},t) = \begin{cases} -\nabla\Phi^2(\rho,T), & if\ p - c_s^2\rho \geq 0 \\ \nabla\Phi^2(\rho,T), & if\ p - c_s^2\rho < 0 \end{cases}, \quad (23)$$

It is obviously that, Eq. (22) is identical with Eq. (23) mathematically. Therefore, each of them can be used as the interparticle interaction in the pseudopotential model based on high-order difference. If there exist any differences resulted from the two interactions, such as reported in the literatures [10, 14], they should stem from the inexactness of the numerical evaluating method.

For the forcing technology, the scheme proposed by Kupershtokh et al. is adopted for incorporating the interaction into the basic lattice Boltzmann equation, because of its accuracy, simplicity, and relax time independence.

As the numerical calculation of the interactions takes an important part in modeling the multiphase flows, how to exactly calculate the gradients in Eqs. (22) and (23) is critical. In order to obtain high accuracies, instead of CDM, HDM should be used to accomplish the computations in the newly proposed model.

## 5. Simulations and results

We now present a series of multiphase flow simulations, in which first-order phase transitions occur, to demonstrate the performances of the pseudopotential multiphase lattice Boltzmann model based on high-order difference. To obtain high numerical stability, the scheme of modifying equation of state proposed by Hu et al. [14], based on the Maxwell construction theory, e.g. $p_{mEOS} = kp_{EOS}$, is applied to all the used equations of state, where $k$ is an adjustable parameter.

### 5.1 Performances of CDM and HDM in real multiphase flows

Hyperbolic tangent functions are usually used as the approximations of the density equilibrium distributions of the multiphase flow systems. However, the real equilibrium distributions of density are not exactly equal to them. A VDW's fluid with relative temperature $Tr = 0.9$ is taken as a test case for investigating the accuracies of different difference methods in real multiphase flows. For the VDW's fluid, in terms of thermodynamic theories, its equation of state is written as

$$p_{EOS}^{VDW} = \rho \psi^{'}(T,\rho) - \psi(T,\rho) = \frac{\rho T}{1-b\rho} - a\rho^2, \tag{24}$$

where $a$ and $b$ are two constant parameters, $\psi(\rho,T)$ is the bulk free energy density of the fluid. The corresponding full pressure tensor can be described as

$$P_{\alpha\beta} = [p_{EOS} - \kappa\rho\nabla^2\rho - \frac{\kappa}{2}(\nabla\rho)^2]\delta_{\alpha\beta} + \kappa\frac{\partial\rho}{\partial x_\alpha}\frac{\partial\rho}{\partial x_\beta}, \tag{25}$$

where $\kappa$ is a constant. When the system is in equilibrium, the following mechanical equilibrium equation must be satisfied

$$\nabla \cdot \ddot{P} = 0 \tag{26}$$

If the interface between two phases is flat and develops along with the y-axis, the fluid densities $\rho(y)$ and its derivatives $\frac{d\rho}{dy}$ can be easily obtained by numerically solving Eq. (26), which will be served as the "exact" values for comparison. On the other hand, by using the "exact" densities $\rho(y)$, we can also calculate its spatial derivatives with CDM or HDM respectively. The calculated results are given in Fig. 4. It displays clearly that, (1) different from Figs. (1) and (2), the "exact" density distributions in equilibrium dos not exactly equal to hyperbolic tangent functions, and their derivatives are not symmetrical about the site of y=0; (2) just the same as for the hyperbolic tangent function in sections 3.1 and 3.2, in general, the calculated derivatives by HDM are much closer to the "exact" values than those by CDM for the real multiphase flows. Thus, in the practical multiphase flow simulations, using HDM for evaluating the gradients in the interactions may obtain much better results.

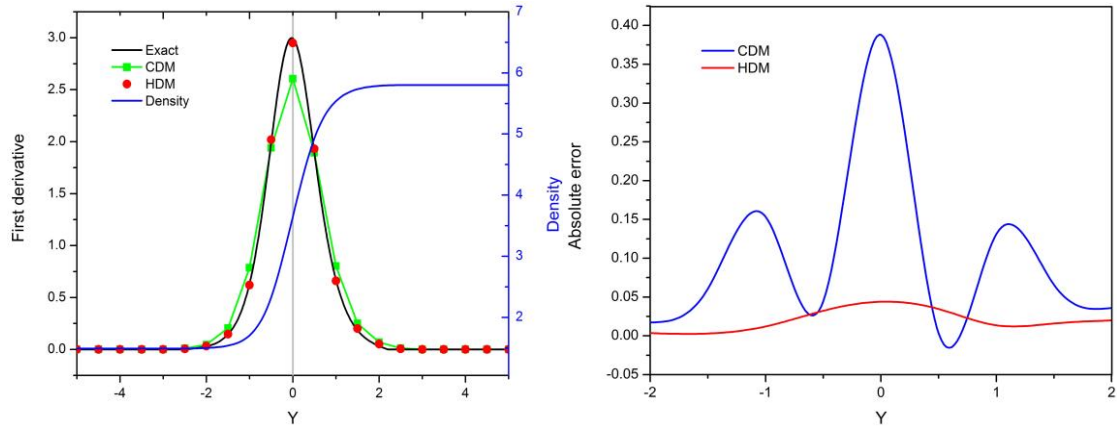

Fig. 4. The derivatives and absolute errors of the equilibrium density distribution of the VDW's fluid with *Tr*=0.9 and h=0.5 by CDM and HDM.

5.2 Coexistence density curves

As the most important thermodynamic consistent tests, the coexistence density curves of the fluid systems with different equations of state are calculated and compared with those predicted by the Maxwell equal-area construction. To systematically investigate the range of application for the proposed model, the tested fluid systems cover the usual equations of state, including VDW EOS, RK EOS, PR EOS and CS EOS.

VDW EOS is the most famous cubic EOS; its formula is described in Eq. (24).

The RK EOS is generally more accurate than the VDW EOS by improving the attraction term

$$p_{EOS}^{RK} = \frac{\rho RT}{(1-b\rho)} - \frac{a\alpha(T)\rho^2}{(1+b\rho)}, \quad (27)$$

where $\alpha(T) = 1/\sqrt{T}$. The Soave modification (RKS) involves a more complicated temperature function, $\alpha(T) = [1+(0.480+1.574\omega-0.176\omega^2)(1-\sqrt{T_r})]^2$, where ω is the acentric factor. The PR EOS is often superior in predicting liquid densities,

$$p_{EOS}^{PR} = \frac{\rho RT}{(1-b\rho)} - \frac{a\alpha(T)\rho^2}{(1+2b\rho-b^2\rho^2)}, \quad (28)$$

where $\alpha(T) = [1+(0.37464+1.54226\omega-0.26992\omega^2)(1-\sqrt{T_r})]^2$ is a factor depending on temperature. The Carnahan-Starling (CS) EOS tends to give better approximations for the repulsive term,

$$p_{EOS}^{CS} = \rho RT \frac{1+b\rho/4+(b\rho/4)^2-(b\rho/4)^3}{(1-(b\rho/4)^3)} - a\rho^2, \quad (29)$$

In the following simulations, the attraction parameter and the volume correction take a = 9/49, b = 2/21 for VDW, PR and RKS EOS, and a = 1, b = 4 for CS EOS. The universal gas constant is R = 1. The acentric factor ω is 0.344 for water (PRW) and 0.011 for methane (PRM). We set the computational domain as 200×200 square with periodical boundary conditions for both horizontal and vertical directions. The relaxation time is taken as 1.6 in general. Considering the interface between two phases is flat in the horizontal plane, the middle part of the domain is initialized as liquid, and the remaining parts are set as gas.

To confirm the performances of the pseudopotential multiphase lattice Boltzmann model based on high-order difference, we simulate the coexistence density curves for different EOSs with both CDM and HDM for Shan-Chen's and Zhang-Chen's interactions respectively. As comparisons, we also perform the simulation with the hybrid interaction proposed by Hu et al. In the simulations, the modifying coefficient *k* of the equations of state takes 0.01 for VDW EOS, 0.02 for RK EOS, and 0.04 for PRW and CS EOSs. The numerical results are drawn on Fig. 5.

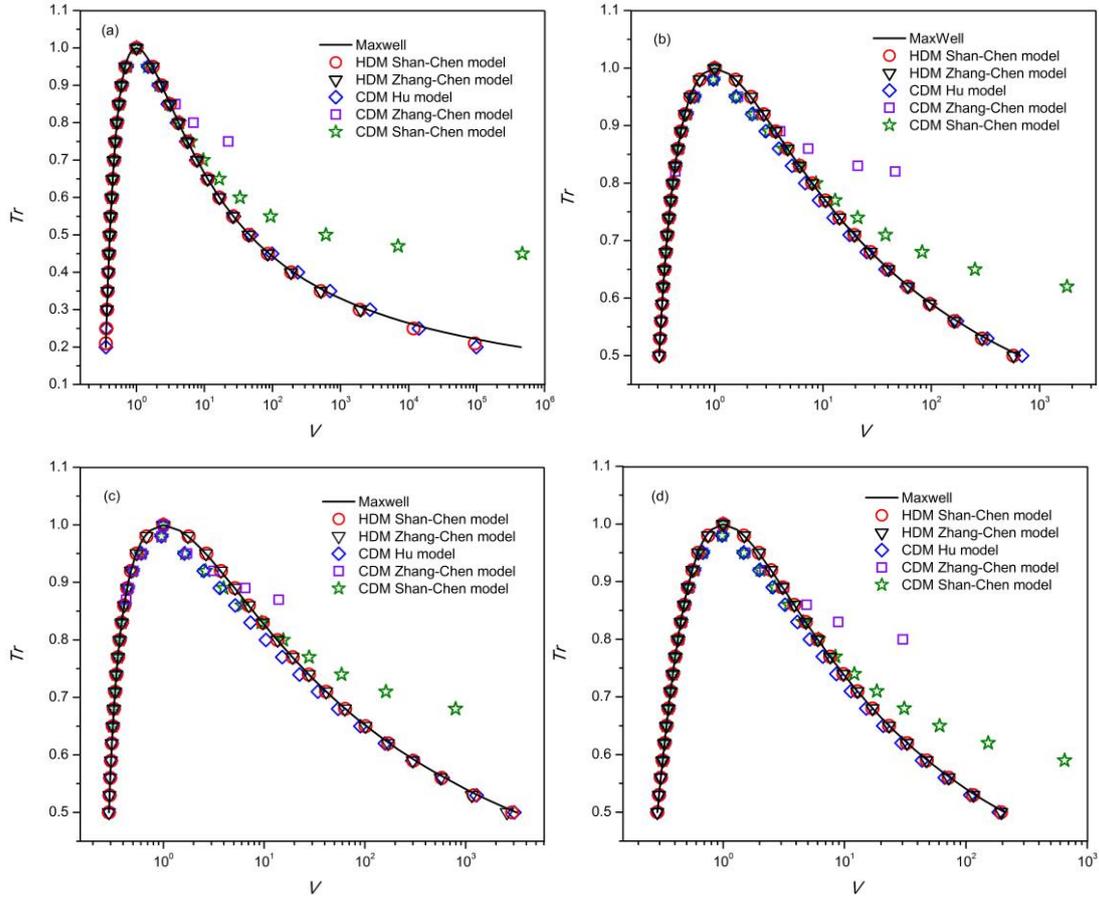

Fig. 5. The coexistence densities of VDW fluid (a), RK fluid (b), PRW fluid (c) and CS fluid (d).

Analyzing Fig. 5 in detail, we can directly draw the conclusions that, (1) just like those reported in the literatures [10, 14], by using the interaction of Shan-Chen or Zhang-Chen, the coexistence densities evaluated with CDM can't be in agreement well with the results predicted by the Maxwell equal-area rule for all selected equations of state; (2) using the hybrid interaction combining both force formulas of Shan-Chen and Zhang-Chen through an adjustable weight coefficient, coexistence densities computed with CDM for the selected equations of state are approximately in agreement of those predicted by the Maxwell equal-area rule，and the agreement is the best for VDW EOS, and worse for the other three EOSs; (3) the coexistence densities with the force formula of either Shan-Chen or Zhang-Chen calculated with HDM are in excellent agreement with the exact values for all of the selected equations of the state, without any interaction combinations and adjustable parameters. As analyzed in section 4, the interaction formulas of Shan-Chen and Zhang-Chen are identical to each other mathematically. It means that, if the numerical calculating accuracy is high enough, they should be identical to each other numerically without any obvious differences. Therefore, these four testing cases illustrate that, firstly the numerical calculation scheme of the interaction takes a key role in the multiphase flow simulations; secondly the usually used CDM may have not enough accuracies to meet the requirements of simulating the multiphase flows; and thirdly HDM have enough accuracies for fully meeting the specific requirements of computing these complex fluid flows. On the other hand, the hybrid interaction scheme of combining the forces of Shan-Chen and Zhang-Chen may not be an excellent solution, as these two forces are identical mathematically and their linear superposition would not generate new physics results. In terms of this view, although there exists an adjustable parameter in the hybrid interaction, the simulated coexistent densities by this hybrid force and CDM can't be equally well in agreement of the "exact" values for different equations of the state. In brief summary, for the pseudopotential multiphase lattice Boltzmann model, CDM may be not an optimal choice as its limited accuracies, and HDM should be an alternative selection due to its reliable accuracies in the real simulations, finally it can be expected that the applications of HDM in pseudopotential multiphase lattice Boltzmann models would provide some more satisfying results.

5.3 Surface tension and Laplace's law

Surface tension is another important property of a multiphase flow system, which is usually also used to verify a multiphase flow model. There are two ways to evaluating the surface tension: firstly, for a system with a flat interface, the surface tension can be evaluated by integrating along with a straight-line perpendicular to the interface, which is so called integration method; secondly, it can be obtained by the famous Laplace's law. A fluid with VDW EOS is chosen for performing the surface tension calculations. At first, we calculate the surface tensions of the fluid at temperatures $Tr = 0.6$, 0.7, and 0.8 by the integration method respectively. Then, we evolve the systems with a droplet in the domain center for different radius and temperatures, and evaluate the surface tensions through Laplace's law. Finally, by comparing the numerical surface tensions from different methods, the proposed model performances for evaluating surface tensions are verified. The results are show in Table 1 and Fig. 6, where $\sigma_S$ and $\sigma_L$ are obtained from the integration method and Laplace's law (so called pressure difference method) respectively and the relative error is defined by $\left|\frac{\sigma_L - \sigma_S}{\sigma_S}\right| \times 100\%$. Table 1 shows that the calculated surface tensions from two different methods are quite close, the largest relative error is 1.58% at temperature $Tr = 0.8$, and as comparisons, these errors are all slightly lower than those reported in the literature [10]. Fitted with the least square method, the calculated temperature dependence of the surface tension is $\gamma \propto (1-T_r)^{1.49}$, which is in excellent agreement with the theoretical result for the VDW EOS $\gamma \propto (1-T_r)^{1.5}$. Fig. 6 displays that, Laplace law is well satisfied for all listed three temperatures, thus the accuracies of the surface tensions calculated with the present model are verified.

Table 1. Surface tensions of VDW fluid with different temperatures

| $T_r$ | 0.4 | 0.5 | 0.6 | 0.7 | 0.8 |
|---|---|---|---|---|---|
| $\sigma_S$ | 0.10885 | 0.08457 | 0.06151 | 0.04044 | 0.02221 |
| $\sigma_L$ | 0.11032 | 0.08544 | 0.06176 | 0.04032 | 0.02186 |
| Relative errors (%) | 1.35 | 1.03 | 0.41 | 0.30 | 1.58 |

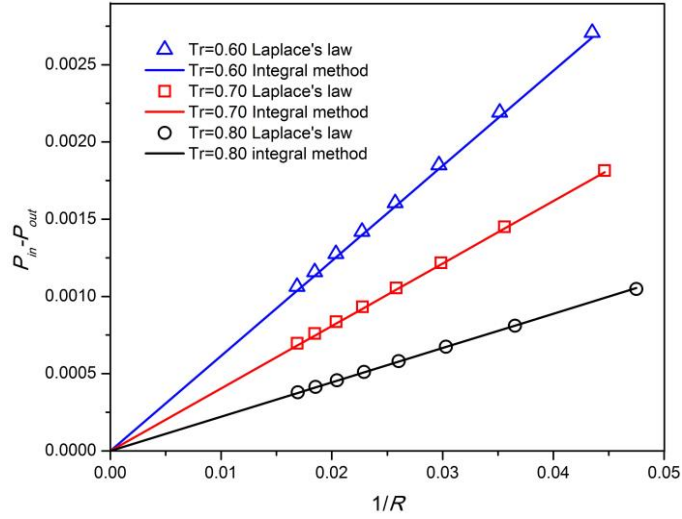

Fig. 6 Pressure differences vs $1/R$ with different temperatures for VDW fluid.

5.4 Spurious currents

Spurious current arising in the simulations of stationary two-phase state is an important property to measure the performances of a multiphase flow model. We investigated the spurious currents of the simulations with different equations of state by the present model. The domains of the simulations are all set to $200 \times 200$ lattice nodes, and a droplet with equilibrium radius 30 lattice nodes was placed in the center. When the simulation reached equilibrium, the maximum fluid velocity near the interface was measured as the spurious velocity. The simulated results are shown in Fig. 7 for VDW EOS, PRW EOS, CS EOS and RK EOS. Fig.7 shows that with the decrease of temperature, the spurious velocities increase apparently for all equations of state, and the spurious velocities are all well below 0.1.

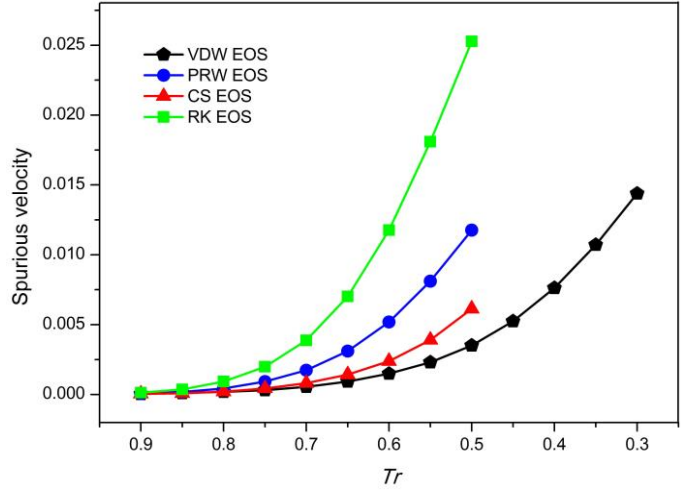

Fig. 7. Spurious velocities in the simulations of VDW, PRW, CS and RK EOSs by the present model and Shan-Chen's interaction.

5.5 Droplet splashing on a thin liquid film

As dynamic testing cases and simple applications, we simulate the droplets with specific initial velocities splashing on thin liquid films.

In the simulations, a two-dimensional planar droplet with PRW EOS is considered, and the domain of the simulations is a rectangle area of $1000 \times 300$ lattice nodes. The liquid film is placed at the bottom of the computational domain, and its height is one tenth of the entire domain height. The radius of the droplet is R = 50, and its impact velocity take U = 0.108 downward along with the y direction. For y-axis direction, the no-slip boundary condition is used, and the periodic condition is adopted for x-axis direction. The reduced temperature is set to be $Tr$ = 0.6, at which the gas/liquid density ratio is close to that of vapor and water.

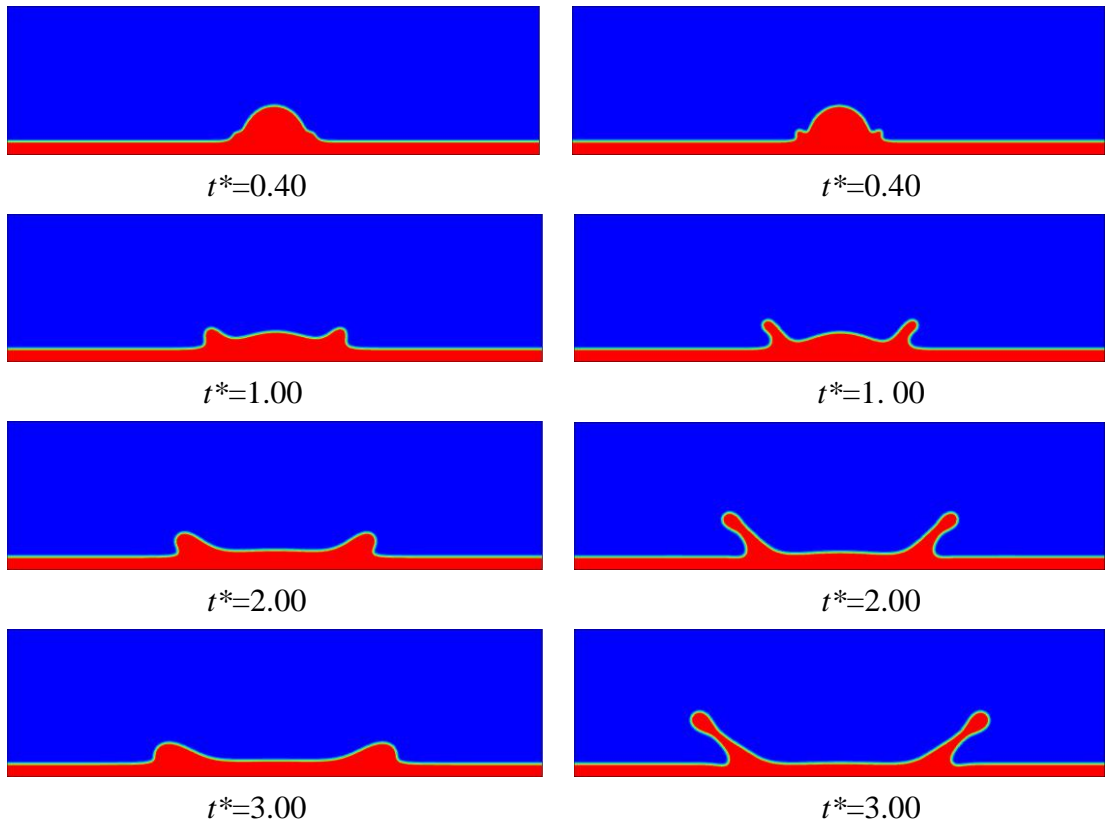

Fig. 8. Snapshots of the impacting process at $Tr = 0.6$, Re = 40 (left) and Re = 100 (right).

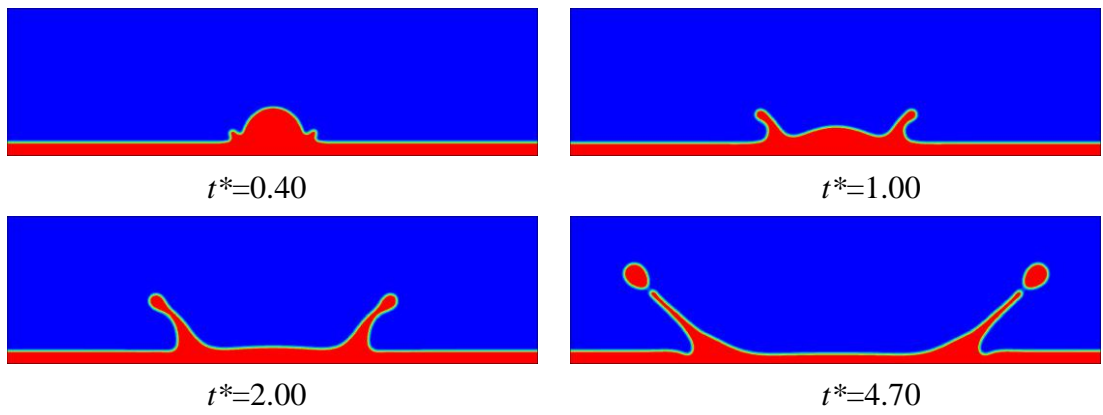

Fig. 9 Snapshots of the impacting process at $Tr = 0.6$ and Re = 173.

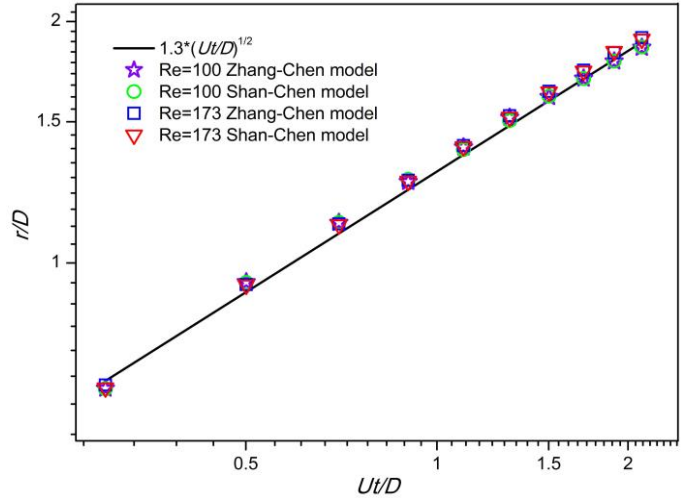

Fig. 10. The predicted spread radius at Re = 100 and 173 as a function of the nondimensional time.

With the above configuration, the simulations for different cases have been done: Re = 40, 100 and 173. The Reynolds number is defined as $Re = UD/\nu_l$, where $D$ is the diameter of the impact droplet. The Weber number is $We = \rho_l U^2 D/\sigma \approx 117$ (the surface tensor $\sigma$ is calculated with the Laplace's law). The snapshots of the impacting process with Re = 40, 100, and 173 are shown in Figs. 8 and 9 respectively, here nondimensional time is defined as $t^* = Ut/D$. Fig. 8 (left) shows that, at Re = 40, the impact of the droplet will not bring about splashing but an outward moving surface wave. With the increase of the Reynolds number, as shown in Fig. 8 (right), when Re =100, a thin liquid sheet will be emitted after the impact of the droplet, which will develop into a crown propagating away from the droplet. When the Reynolds number is big enough, such as Re = 173 (Fig. 9), a thinner liquid sheet will be formed at the intersection between the droplet and the liquid layer. Then the sheet tilts upward and evolves into an almost vertical lamella whose end rim is unstable and will eventually break up into secondary droplets. This well-known phenomenon of droplet splashing can be seen clearly in Fig. 9 at $t^* = 4.7$. Another important property related with droplet splashing on a thin liquid film is that the spread radius r generally obeys the power law $r/D \approx C\sqrt{Ut/D}$ at short times after the impact, in which the constant *C* is usually greater than 1.1 for two-dimensional modeling of droplet splashing [28]. We also investigated this problem by the present model with both interactions of Shan-Chen and Zhang-Chen, and the results are given in Fig. 10. The present numerical results for the spread radius are in overall agreement with the prediction of the power law $r/D \approx 1.3\sqrt{Ut/D}$, and again the identical results of both two interactions are obtained in the dynamic situations with HDM.

## 6. Conclusion

By analyzing the numerical accuracies of the derivatives of hyperbolic tangent functions, we find that the usually used CDM may introduce non-negligible errors, which might be large enough to bring about incorrect results, and HDM gives excellent accuracies, which are much better than those of CDM. Then a pseudopotential multiphase lattice Boltzmann model based on high-order difference is proposed. To test the performances of the present model, numbers of numerical simulations are conducted, and the following conclusions are drawn.

1. In the real phase transition situations of the fluid with VDW EOS, the usually used CDM may introduce significant errors for the equilibrium density gradients, which might lead to incorrect results, however HDM gives excellent results (almost equal to the exact values).
2. By using CDM, neither the interactions of Shan-Chen nor Zhang-Chen can well depict the phase transition behaviors of any fluids with the usually used equations of state.
3. Using the hybrid interaction with an adjustable parameter through CDM, the behaviors of the fluids with all selected equations of state can be approximately described with different accuracies, and the results of VDW EOS are better than for the other three EOSs.
4. The pseudopotential multiphase lattice Boltzmann model based on high-order

difference needs not any combinations of the interactions and any adjustable interaction parameters. By using this model, the simulated phase transitions for the usually used equations of state are all in excellent agreement with those predicted by Maxwell equal-area rule.

5. Whether using Shan-Chen's or Zhang-Chen's interaction in the present model, all the simulated coexistence density curves for the fluids with the selected equations of state are equally in well agreement with those predicted by Maxwell equal-area rule. This supports our analysis that the two interactions are equal mathematically and indicates that it is the numerical errors from CDM to make them different as shown in the literatures [10, 14].

6. Owing to its high accuracies, the spurious currents, for all the selected equations of state are obviously reduced and all well below 0.1 in the present model. With this new model, the stability is obviously enhanced at a wider temperature range.

7. With the present model, the dynamics of droplet splashing is correctly reproduced, and the predicted spread radius is found to well obey the power law reported in the literature. This may illustrate that the newly proposed model can be easily and reliably applied to various practical simulations.

As its high accuracies and easy implements in the multiphase flow simulations, the pseudopotential multiphase lattice Boltzmann model based on high-order difference can be expected to be applied on more general complex fluid systems and obtain some more exciting results.

**Acknowledgement**

This work was supported by the National Natural Science Foundation of China (Grants Nos. 11462003 and 11362003), Guangxi Science and Technology Foundation of College and University (Grant Nos. KY2015ZD017 and KY2016YB063), Guangxi "Bagui Scholar" Teams for Innovation and Research Project, Guangxi Collaborative Innovation Center of Multi-Source Information Integration and Intelligent Processing.